\documentclass{aa}
\usepackage{graphicx,natbib}
\bibpunct{(}{)}{,}{a}{}{,}
\begin{document}

   \title{Phase mixing of a three dimensional magnetohydrodynamic pulse}

   \author{D. Tsiklauri, V.M. Nakariakov and G. Rowlands}

   \offprints{David Tsiklauri, \\ \email{tsikd@astro.warwick.ac.uk}}

   \institute{Physics Department, University of Warwick, Coventry,
   CV4 7AL, England }

   \date{Received 4 October 2002 / Accepted 16 December 2002}

\abstract{Phase mixing of a three dimensional magnetohydrodynamic (MHD) pulse
is studied in the compressive, three-dimensional (without an ignorable
coordinate) regime.  It is shown  that
the efficiency of decay of an Alfv\'enic part 
of a compressible MHD pulse is 
related linearly to the degree of  localization of
the pulse in the homogeneous transverse direction.
In the developed stage of phase mixing (for large times), 
coupling to its compressive part
does not alter the power-law  decay 
of an Alfv\'enic part of a compressible MHD pulse.
The same applies to the dependence upon the resistivity 
of the Alfv\'enic part of the pulse.
All this implies that the dynamics of Alfv\'en waves can
still be qualitatively understood in terms of the previous 2.5D
models. Thus, the phase mixing remains a relevant paradigm 
for the coronal heating 
applications {\it in the realistic 3D geometry and compressive plasma}.
 }
\titlerunning{Phase mixing of a 3D MHD Pulse ...}
\authorrunning{Tsiklauri, Nakariakov \& Rowlands}
\maketitle

\section{Introduction}

The phase mixing of {\it incompressible}
Alfv\'en waves \citep{hp83} has been intensively studied
as a possible mechanism of coronal heating in open magnetic
structures of the polar regions of the Sun.
Also, there are recent observational indications
that the same mechanism operates in coronal loops \cite{oa02}.
In a {\it compressible} plasma,
phase mixing of linearly polarized plane
Alfv\'en waves may provide  enhanced heating through the 
nonlinear generation of fast magnetoacoustic waves. This phenomenon
has been intensively studied by \citet{malara,nrm97,nrm98,Botha,td1,td2}
both analytically and numerically.

There are three main motivations for further development of 
this work: The first, as  was recently shown by 
\citet{hbw02}, is that  
phase mixing of {\it localized} single Alfv\'en
pulses results in a slower, power-law damping as opposed to the
standard $\propto \exp(-t^3)$ one for  harmonic Alfv\'en waves 
\citep{hp83}. This
suggests that localized Alfv\'enic perturbations can transport
energy higher into the corona than  harmonic ones. 
Also, \citet{hbw02b} showed that this power-law damping
is replaced by even faster power-law decay as the number 
of pulses is increased. Such
perturbations could be generated e.g. by transient events such as
solar flares, coronal mass ejections, etc \citep{r01}.
The second follows from the recent demonstration by \citet{td3} that
in the linear regime and in 3D geometry, generation of 
compressive perturbations
significantly modifies the dynamics 
of the Alfv\'enic part of a MHD pulse. In fact, 
while solving the propagation part of the problem, i.e. considering
an ideal plasma limit, it has been
shown that an initially localized 3D Alfv\'en pulse
interacting with a 1D transverse inhomogeneity evolves
to a MHD pulse {\it with a significant compressible component}.
This 3D MHD pulse is subject to phase mixing.
Therefore, as a natural step forward,
in this study we include a finite 
plasma resistivity in order to investigate {\it quantitatively} 
dissipation of a 3D MHD pulse via phase mixing.
The third follows from Parker's (1991) 
earlier suggestion that in order for the phase mixing
mechanism to work, it requires the presence of a third ignorable
coordinate. Although it has been justly observed that
having an ignorable coordinate seems unlikely in a filamentary
corona, and, also,
in the absence of an ignorable coordinate, the \lq\lq Alfv\'en-type
waves" are coupled to the \lq\lq fast magnetosonic-type modes", 
we believe it was not
 justified to dismiss the importance of
 phase mixing without a rigorous, 
fully three dimensional study.
\section{The model}
In our model we use the MHD equations 
\begin{equation}
\rho \; {\partial_ t} {\vec V} +
\rho(\vec V \cdot \nabla) \vec V = - \nabla p -(4 \pi)^{-1}
\vec B \times {\rm curl} \vec B, \label{1}
\end{equation}
\begin{equation}
{\partial_ t}{\vec B}= {\rm curl} (\vec V \times \vec B)+
\eta \Delta {\vec B},
\label{2}
\end{equation}
\begin{equation}
{\partial_ t} p + \vec V \cdot \nabla p + \gamma p
\nabla \cdot \vec V=(\gamma -1) \eta ({\rm curl} \vec B)^2,
\label{3}
\end{equation}
where $\vec B$ is the magnetic field, $\vec V$ is the plasma velocity,
$\rho$ is the plasma mass density, and $p$ is the plasma thermal pressure.
In what follows we use $5/3$ for the value of $\gamma$.
Also, here $\eta$ denotes plasma resistivity.

We solve equations (\ref{1})-(\ref{3}) in Cartesian 
coordinates ($x,y,z$).
Note that we solve a fully 3D problem retaining variation in the
$y$-direction, i.e. ($\partial / \partial y \not = 0$). 
A uniform magnetic field $B_0$ is in the $z$-direction.
The plasma configuration 
has a one-dimensional inhomogeneity in the equilibrium density
$\rho_0(x)$ and temperature $T_0(x)$, while
the unperturbed thermal pressure, $p_0$, is taken to be constant
everywhere.
\begin{figure}[]
\resizebox{\hsize}{!}{\includegraphics{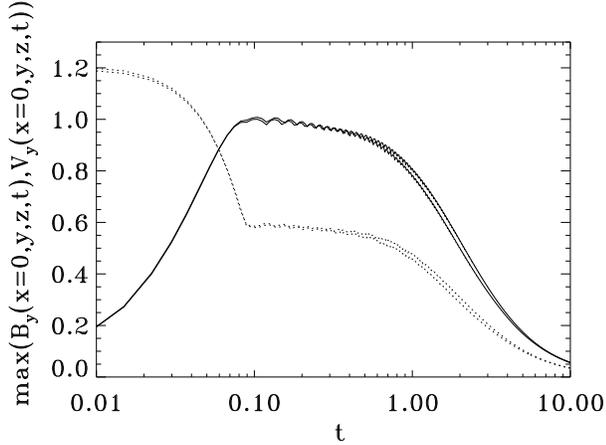}} 
\caption{Time evolution of 
$max(B_y(x=0,y,z,t)$ (solid lines) and 
$max(V_y(x=0,y,z,t)$ (dotted lines). Note the plot is log-normal. 
Here, $\alpha_y=1.0$. 
The thick lines represent simulation results with the doubled 
resolution in $y$-direction. Note progressive convergence of the
solutions with the increase of simulation time.} 
\label{fig1}
\end{figure}
\begin{figure}[]
\resizebox{\hsize}{!}{\includegraphics{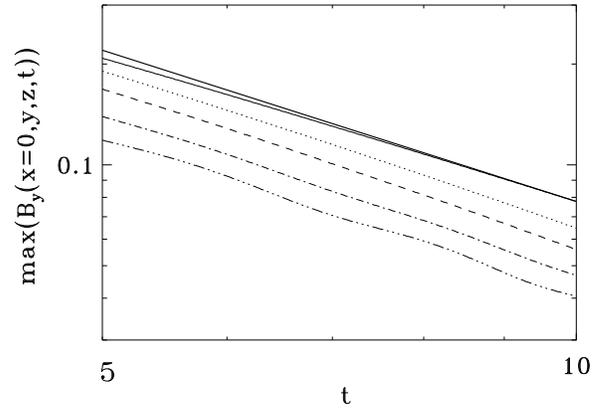}} 
\caption{The developed stage of phase mixing: time evolution of 
$max(B_y(x=0,y,z,t)$ for large times (log-log plot). 
The thick solid line represents the analytic solution (Eq.(\ref{9}) with an
additional pre-factor 0.5, to account for the
split of the initial pulse into two D'Alambert solutions), while the 
thin line is our numerical simulation result (for $\alpha_y=0$).
Note that the deviation between the two progressively vanishes.
The dotted, dashed, dash-dotted and dash-triple-dotted curves
correspond to the numerical solutions for 
$\alpha_y=0.5, \; 1.0, \; 1.5, \; 2.0$ respectively.} 
\label{fig2}
\end{figure}
\section{Numerical Results}
Evolution of a purely Alfv\'enic perturbation 
allows a full analytical treatment and this is presented
in Appendix A. However,
if an initial Alfv\'enic perturbation  
depends on the $y$-coordinate, contrary to the case
studied in the Appendix A, the
linearized version of the system of equations (\ref{1})-(\ref{3})
no longer allows decoupling of Alfv\'enic and 
compressive perturbations.
In fact, all three modes, Alfv\'en, fast and slow
magnetosonic,  are inter-coupled. Thus, they cannot be separated.
Because of the complexity of the problem in this case,
we resort to direct numerical simulations of Eqs.(\ref{1})-(\ref{3})
using the numerical code {\it Lare3d} \citep{Arber}.
We employ the following initial conditions: 
\begin{equation}
V_y|_{t=0}=\cases{-A C_A(x)\left[1+\cos(10 \pi z)\right] &$$\cr
\times \exp\left[-\alpha_y^2\left(y+
\frac{\displaystyle 1}{\displaystyle 
\sqrt{2}\alpha_y}\right)^2 \right] &$|z|\leq 0.1$ ,\cr
0 &elsewhere \cr}
\label{10}
\end{equation}
The rest of the perturbations at $t=0$ are
set to zero (including $B_y$, in this way we guarantee 
fulfillment of  div${\vec B}=0$, at $t=0$). 
In Eq.(\ref{10}) $A$ is the amplitude of the
wave, $C_A(x)$ is the local Alfv\'en speed (see Appendix A)
and $\alpha_y$ is a free parameter that controls
the strength of the gradient in the $y$ direction of the initial
perturbation. In effect, $\alpha_y$ controls 
the strength of the coupling between the Alfv\'enic and
compressive (fast and slow magnetosonic) components of the
MHD pulse.
This particular choice of $y$-dependence in Eq.(\ref{10}) is
dictated by a desire to have the maximum of the absolute value 
of $\partial_y V_y$, which is 
responsible for the coupling to the compressive modes,
at the middle of the simulation cube, where the background
plasma inhomogeneity is the strongest.
Since {\it Lare3d} is a fully non-linear numerical code, we simply
set small (linear) initial amplitude of the perturbation (given by Eq.(\ref{10}))
as $A=0.0001$, while the
output results are normalized to $A$.
The simulation cube size was set by the limits $-11.0 \leq x,y,z \leq 11.0$.
Boundary conditions used in all our simulations are
zero-gradient in all three spatial dimensions.
In fact, our strategy was to forbid any
generated wave fronts to reach the boundaries to avoid
contamination of the solution. This of course severely
restricts simulation end time. However, with 
the computational resources at our disposal, it was still
possible to reach the developed stage of phase mixing.
In all our numerical simulations we fix the
plasma $\beta$ at $10^{-3}$ (which is a plausible value for solar corona) 
and dimensionless resistivity (see Appendix A), $\eta$,
at $5 \times 10^{-4}$ (which is the same value used by \citet{hbw02})
unless stated otherwise.
\begin{figure}[]
\resizebox{\hsize}{!}{\includegraphics{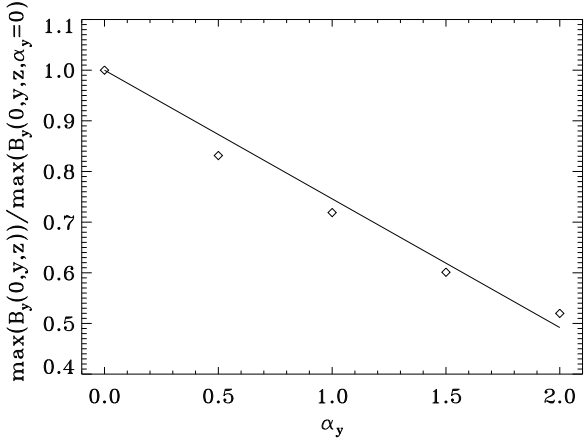}} 
\caption{Efficiency of decay of the Alfv\'enic part of 
MHD pulse, $max(B_y(x=0,y,z)/max(B_y(x=0,y,z,\alpha_y=0)$,
as a function of coupling to the compressive
modes strength parameter, $\alpha_y$ at $t=10$.}
\label{fig3}
\end{figure}
\begin{figure}[]
\resizebox{\hsize}{!}{\includegraphics{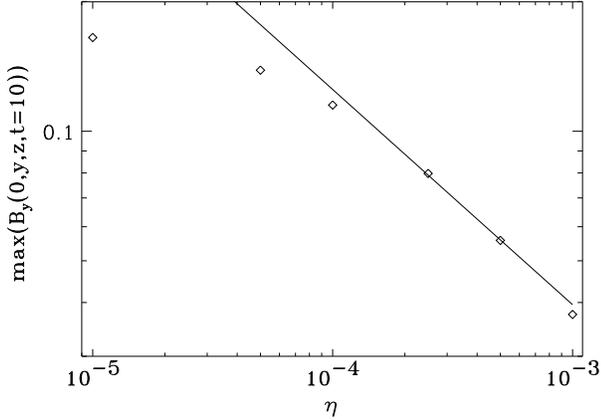}} 
\caption{Dependence  of 
$max(B_y(x=0,y,z,t=10)$ upon resistivity $\eta$. 
The solid line represents the analytic solution (Eq.(\ref{9}) with an
additional pre-factor 0.5, to account for the
split of the initial pulse into two D'Alambert solutions and
a further pre-factor 0.7188, which comes from the 
middle point of Fig.~(\ref{3})), while open
diamonds are numerical simulation results.
Here, $\alpha_y=1.0$.}
\label{fig4}
\end{figure}
In Fig.~(\ref{fig1}) we track the evolution of the maximal value 
(over the whole simulation cube at a given time) of
$B_y$ and $V_y$ at the middle of the density inhomogeneity at $x=0$, where
phase mixing significantly enhances the dissipation.
Note that $max(|V_y(x=0,y,z,t)|)$ at $t=0$ is not
unity as it is normalized to $A$, not to $max(|V_y(x=0,y,z,t=0)|)$.
We gather from this graph that although $B_y$ was initially
zero, it is rapidly generated (on the time
scale of $\sim 0.05$, which is the ratio of length-scale of the
pulse $1/10 \pi$ to  $C_A(0)=0.6$), and  subsequently it decays
due to phase mixing.
Most of our numerical 
runs were done with $200 \times 72 \times 400$ resolution
(with a non-uniform grid), and we checked
convergence of our results by doubling the number of 
grid points in the $y$-direction.
Coincidence of the two runs was satisfactory and, in fact, it progressively
improves with the increase of the simulation time.
Although we were unable to double the resolution in $x$ and $z$,
proof of the convergence of our results comes from the fact 
(see Fig.~(\ref{fig2})) that the
analytic result (given by Eq.(\ref{9}) with an
additional pre-factor 0.5 to account for the splitting of initial
$V_y$ perturbation into two D'Alambert solutions \citep{td3}) and the numerical
solution in the case $\alpha_y=0$ (when there is a complete decoupling
of the Alfv\'en and the compressive modes) coincide perfectly for
large $t$s. In fact, at $t=9.998143$, the absolute deviation
between the two is
$(0.077836-0.077785)/0.077785=0.0006$, 
which is a remarkably good match for a fully three
dimensional simulation. Figure~(\ref{fig2}) demonstrates that: 
(A) an increase in the coupling strength
parameter,  $\alpha_y$, results in an overall decrease of the
Alfv\'enic part ($B_y$) of the MHD pulse, as expected, and 
(B) At large times, all lines on the log-log plot tend to become
parallel. Therefore, the amplitude of the 
Alfv\'enic component of the MHD pulse decays with the same
power-law ($B_y \propto t^{-3/2}$) as in the case of a
localized, {\it pure}, Alfv\'en wave (which occurs only 
when $\alpha_y=0$), 
i.e. {\it coupling to the compressive waves does not affect
the decay power-law of the Alfv\'enic part of the MHD pulse}.
In order to investigate and quantify point (A) further,
in Fig.~(\ref{fig3}) we plot  the efficiency of decay of the Alfv\'enic part of 
the MHD pulse, which we define as 
$max(B_y(x=0,y,z)/max(B_y(x=0,y,z,\alpha_y=0)$,
as a function of coupling to the compressive
modes strength parameter, $\alpha_y$ at $t=10$. Open diamonds represent
numerical simulation results, while a straight line is the best fit,
$1.0-0.25403 \alpha_y$. We conclude that the efficiency of decay
depends {\it linearly} 
upon the degree of  localization of
the pulse in the homogeneous transverse direction, i.e.
the coupling strength parameter, $\alpha_y$.
As far as point (B) is concerned, apart from the visual manifestation that
at large times all lines tend to become parallel to each other, an actual
fitting procedure also confirms that, indeed, amplitude of the 
Alfv\'enic part of the MHD pulse decays as a
power-law ($ \propto t^{-3/2}$) within a few percent error 
to the power-law index.
This is not a trivial result. From this we reach the 
conclusion that {\it the 
dynamics (phase mixing) of weakly non-plane Alfv\'en waves can
still be qualitatively understood in terms of the previous 2.5D
models}. 
A certain level of understanding of this profound conclusion
can be achieved by analyzing 
a linearized version of the system of 
equations (\ref{1})-(\ref{3}) for $\beta =0$.
In this case, coupling between the Alfv\'en ($B_y$) and fast magnetosonic
mode ($B_x$) (the slow mode is absent in this approximation) is 
described by the
following system of equations:
\begin{eqnarray}
\nonumber
\left[ \partial^2_{tt} -C_A^2(x) (\partial^2_{xx}+
\partial^2_{zz})\right] B_x  = & & \\
\eta \partial_{t} \partial^2_{xx}B_x + & C_A^2(x)\partial^2_{xy}B_y & 
\label{11}
\end{eqnarray}
\begin{eqnarray}
\nonumber
\left[ \partial^2_{tt} -C_A^2(x) (\partial^2_{yy}+
\partial^2_{zz})\right] B_y  = & & \\
\eta \partial_{t} \partial^2_{xx}B_y + & C_A^2(x)\partial^2_{xy}B_x & 
\label{12}
\end{eqnarray}
From Eqs.(\ref{11}) and (\ref{12}) it follows that the terms
responsible for the 
coupling between the Alfv\'en and fast magnetosonic
mode are $C_A^2B_{x,y}/(L_x L_y)$ (where $L_{x,y}$ are spatial scales
in $x$ and $y$), 
while the terms responsible for the
dissipation are $\eta B_{x,y}/(t_AL_x^2)$ (where $t_A$ is a 
typical Alfv\'en time scale).
At large $t$s, because of the action of phase mixing  
in the region of density inhomogeneity,
dissipative terms ($\propto L_x^{-2}$) are much greater
than the coupling terms ($\propto L_x^{-1}$). Therefore,
{\it for $t \gg 1$ the dynamics of $B_y$ is governed by the dissipation
rather than its coupling to the compressive fast mode}.
This, serves as an explanation why,
at large times, all lines in Fig.~(\ref{fig2}) tend to become
parallel, i.e. the amplitude of 
Alfv\'enic component of the MHD pulse decays as the same
power-law ($B_y \propto t^{-3/2}$) as in the case of
localized, pure, Alfv\'en wave.

\begin{figure}[]
\resizebox{\hsize}{!}{\includegraphics{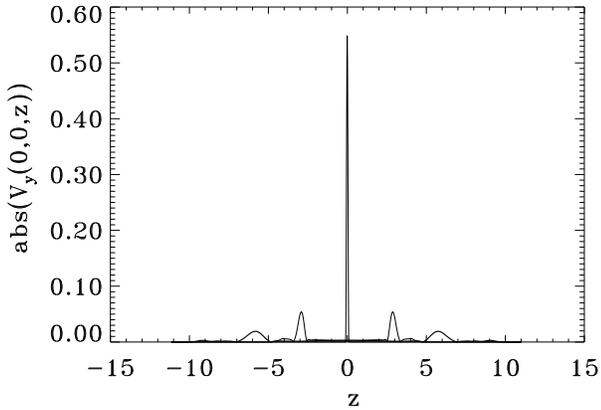}} 
\caption{Evolution  of 
$|V_y(x=0,y=0,z)|$ $z$-profile in time. 
Thin line (spike in the middle) is at $t=0$, 
while thick and the thickest lines depict $|V_y(x=0,y=0,z)|$ at
$t=5.0$ and 10.0. Here, $\alpha_y=1.0$.}
\label{fig5}
\end{figure}

Also, we investigated the dependence of the Alfv\'enic part of the 
MHD pulse, which we quantify by $max(B_y(x=0,y,z,t=10)$,
upon resistivity, $\eta$. 
The results are presented in Fig.~\ref{fig4}.
On the one hand, an accurate analytic solution
(\ref{9}) for $B_y$, which scales as $\propto 1/\sqrt{\eta}$, 
was obtained for the case of $\alpha_y=0$. 
On the other hand, point (B), mentioned above, indicates that
when $\alpha_y \not = 0$, the solution (\ref{9})
should be altered by a further pre-factor that can be
determined e.g. from Fig.~(\ref{3}) (in this case, as $\alpha_y=1$, it is
0.7188). Therefore, we can see that this way constructed
analytic solution gives a tolerably good fit in Fig.~(\ref{fig4}).
The deviation in the small $\eta$ end can be easily explained
by the fact that the solution (\ref{9}) was derived in 
the developed phase mixing approximation, which for such
small values of $\eta$ would require numerical runs with 
end-simulation times much larger than $t=10$ (the latter is not
possible with current computational facilities, as it would require
further increase of the simulation box size).
Thus, we conclude that the coupling to its compressive part
does not alter the $1/\sqrt{\eta}$ behavior
of the Alfv\'enic part of a compressible MHD pulse.

In order to fully understand the dynamics of 
the Alfv\'enic part of the MHD pulse under our initial conditions
as well as to appreciate fully the extent of the efficiency of the 
phase mixing for the coronal plasma heating, in Fig.~(\ref{fig5}) we show the
evolution  of the absolute value of the $z$-profile in the middle
of the plasma inhomogeneity of $V_y$  with time.
We gather from this graph that the initial, very localized,
profile splits into two D'Alambert solutions 
propagating in opposite directions along the field
(cf. \citet{td3}), which then
dissipate rapidly towards a diffuse, Gaussian shape.

Finally, we would like to reiterate the importance of
plasma compressibility on the MHD wave dynamics in the context
of wave-based theories of coronal heating 
and MHD turbulence (cf. \citet{td3}).
Our present numerical simulations confirm that
when $\alpha_y \not =0$, coupling to the Alfv\'enic part of the 
MHD pulse quite efficiently generates a density perturbation
(which was initially zero) on a sizeable fraction of the initial
Alfv\'en amplitude.
\section{Conclusions}
In summary, we have established that:
\begin{itemize}
\item Efficiency of the decay of the Alfv\'enic part 
of a compressible MHD pulse is 
related linearly to the 
degree of  localization of
the pulse in the homogeneous transverse direction, i.e.
the coupling strength parameter $\alpha_y$.
\item In the developed stage of phase mixing (for large times),
coupling to its compressive part
does not alter the $1/\sqrt{\eta}$ behavior
of the Alfv\'enic part of a compressible MHD pulse.
\item The dynamics of Alfv\'en waves can
still be qualitatively understood in terms of the previous 2.5D
models, e.g., \citet{hbw02}'s law ($B_y \propto t^{-3/2}$) for spatially
localized, single pulses, 
survives in the case of 3D compressive MHD waves (in the absence of
an ignorable coordinate). 
\item  All of the above implies that 
phase mixing remains a relevant paradigm for the coronal heating 
applications {\it in realistic 3D geometry and compressive plasma}.
\end{itemize}

 It is useful to give estimates of typical scales
of the problem. We choose a typical (normalization) 
length scale of the transverse inhomogeneity L=10 Mm and 
Alfv\'en speed $C_A=1000$ km s$^{-1}$, which are
appropriate for example for coronal plumes.
The coronal value for shear (Braginskii) viscosity, 
by which Alfv\'en waves damp, 
is estimated to be about $\eta=1$ m$^{2}$ s$^{-1}$ (e.g. \citet{hbw02}, 
which translates into
$10^{-10}$ in the dimensionless units).
However, as some recent studies suggest \citep{netal99,oa02}
the actual value of the shear viscosity
is possibly much larger due to micro-turbulence, 
e.g. $10^{9.2 \pm 3.5}$ m$^{2}$ s$^{-1}$ \citep{oa02}.
Effectively, this implies 
that the Alfv\'en waves  damp by bulk Braginskii viscosity
rather than the shear one.
The latter estimate is based on the 
observations of
standing (kink) Alfv\'en waves in the coronal loops. Thus,
its use in our estimates is justified provided
the propagating Alfv\'en waves, studied here, 
are affected by the same dissipation mechanism as the
standing ones.
Using Eq.(\ref{8}) (for $x=0$) we can estimate the characteristic time of
damping of the amplitude of an Alfv\'en wave (when it reduces, say,
$e$-times)
due to phase mixing. 
Thus, we obtain approximately $2745.22$ s for 
$\eta=1$ m$^{2}$ s$^{-1}$ and $0.16 - 34.56$ s for 
$\eta=10^{9.2 \pm 3.5}$ m$^{2}$ s$^{-1}$.
This translates into 1647.13 Mm (the Alfv\'en speed in the middle of
inhomogeneity is 600 km s$^{-1}$) for the damping length, 
for $\eta=1$ m$^{2}$ s$^{-1}$, and $0.10 - 20.74$ Mm 
for $\eta=10^{9.2 \pm 3.5}$ m$^{2}$ s$^{-1}$.
A similar analysis for the harmonic waves (now using Eq.(\ref{7})
and substituting $k=2 \pi /  \sigma$) yields 
a damping length of 609.47 Mm for $\eta=1$ m$^{2}$ s$^{-1}$ and 
$0.04 - 7.67$ Mm for $\eta=10^{9.2 \pm 3.5}$ m$^{2}$ s$^{-1}$.
These estimates would be affected if we choose
different (probably higher) Alfv\'en speeds or
if the pulse is wider.
Also, note that in our numerical runs, the width of the Gaussian
Alfv\'en pulse in the longitudinal ($z$) direction was fixed at
$ \sigma \approx 0.03$,
i.e. 0.3 Mm, while in the transverse ($y$) direction the width,
$1 / \sqrt{2} \alpha_y$,  
was varied from $+\infty$ down to 0.35, i.e.
from $+\infty$ down to 3.5 Mm.

\begin{acknowledgements}
DT acknowledges financial support from PPARC.
Numerical calculations of this work were
performed using the PPARC funded Compaq MHD Cluster at St Andrews.
GR acknowledges financial support from a Leverhulme Emeritus
Fellowship.
\end{acknowledgements}

\appendix
\section{}
Here, we present an alternative, more rigorous, way to derive
the power-law decay of $B_y$, suggested by \citep{hbw02}.
If the initial Alfv\'enic perturbation (perturbing $V_y$ and $B_y$) 
does not depend on the $y$-coordinate,
the linearized version of the system of equations (\ref{1})-(\ref{3})
allows for a 
complete decoupling of Alfv\'enic and compressive perturbations.
Therefore, in order to investigate phase mixing
quantitatively, we solve the 
following initial value problem with zero-gradient boundary
conditions:
\begin{equation}
\left[ \partial^2_{tt} -C_A(x)^2 \partial^2_{zz}\right] B_y=\eta \partial_{t}
\partial^2_{xx}B_y,
\label{4}
\end{equation}
\begin{equation}
B_y|_{t=0}=f(z)
\label{gen}
\end{equation}
Here, $C_A(x)$ is a dimensionless
Alfv\'en speed.
In our numerical simulations we have used the following
profile for  $C_A(x)$:
$C_A(x)=1 $ for $x < -0.5$, 
$C_A(x)=0.6+0.4 \cos[ \pi (x+0.5)]$ for $x \in [-0.5,0.5]$, and
$C_A(x)=0.2 $ for $x > 0.5$ (we employ here the same normalization as in
\citet{td3} and,  $\eta$ is normalized to the product
of reference length and reference Alfv\'en speed as in \cite{hbw02}).
Note, that as in the developed stage of phase mixing (for large times)
$\partial^2_{xx} \gg \partial^2_{yy}, \partial^2_{zz}$,
in the forthcoming analytic treatment, 
we only retain largest (over $x$-coordinate) derivatives
in the Laplacian (while for generality we still 
keep the full Laplacian in our numerical simulations, see below).
Next, we introduce the following Lagrangian coordinates and
slow (dissipation) time scale:
$\bar x=x$, $\xi=z-C_A(x) t$, $\tau = \varepsilon t$, with
$\varepsilon \ll 1$. 
In these variables, the leading term on the left hand side of Eq.(\ref{4})
is $-2 \varepsilon C_A(x) \partial^2_{\tau \xi} B_y$, while on the
right hand side the leading term is 
$- \eta C_A(x) \partial_{\xi} [ C^\prime_A(x)^2 
(\tau / \varepsilon)^2 \partial^2_{\xi \xi} B_y]$.
Here, prime denotes a derivative over $x$.
After integration over $\xi$ and introduction of yet another
auxiliary variable, $s= \eta C^\prime_A(x)^2 \tau^3/ (6 \varepsilon^3)=
\eta C^\prime_A(x)^2 t^3/ 6$, we obtain the following 
(diffusion) equation for $B_y$: 
\begin{equation}
\partial_s B_y= \partial^2_{\xi \xi} B_y,
\label{diff_e}
\end{equation}
which can be easily
integrated:
\begin{equation}
B_y={\frac {1} {2 \sqrt{\pi s}}} \int_{-\infty}^{+\infty}
\exp \left[{-{\frac{(\xi-\xi')^2}{4 s}}}\right] 
\, B_y(\xi',t=0) \, d \, \xi'.
\label{6}
\end{equation}
From Eq.(\ref{6}) it immediately follows that
in the developed stage of phase mixing, i.e. when $t,s \to +\infty$,
the solution reduces to 
\begin{equation}
B_y={\frac {1} {2 \sqrt{\pi s}}} \int_{-\infty}^{+\infty}
B_y(\xi',t=0) \, d \, \xi'.
\label{6inf}
\end{equation}
Therefore, provided the integral
\begin{equation}
\int_{-\infty}^{+\infty}
B_y(\xi',t=0) \, d \, \xi',
\label{byinf}
\end{equation}
is finite (which is not always so e.g. for the harmonic initial conditions,
see below), $B_y$ scales as $1/\sqrt{s}$ or equivalently
$B_y \propto \eta^{-1/2}  t^{-3/2}$.

If we substitute a harmonic wave with initial conditions
$B_y(\xi',t=0)=\exp(i k \xi')$ into Eq.(\ref{6}), simple
integration yields
\begin{equation}
B_y= e^{- \eta C^\prime_A(x)^2 t^3 k^2/ 6} \, e^{-i k(z-C_A(x) t)},
\label{7}
\end{equation}
thus, we recover the well-known Heyvaerts \& Priest's solution.
For the spatially localized (in $z$) initial conditions  
\begin{equation}
B_y|_{t=0}= 
\biggl\{ \begin{array}{ll}
 1+\cos(10 \pi z)& 
z \in [-0.1,0.1], \\
 0& {\rm elsewhere}  
\end{array}
\label{5}
\end{equation}
following \citet{hbw02}
we expand $B_y(\xi',t=0)$ in terms of Hermite polynomials
$B_y(\xi',t=0)=\sum_{n=0}^{\infty} 
\alpha_n e^{-{\xi'}^2/2 \sigma^2} {\cal H}_n({\xi'}/ \sigma)$ retaining
only the fundamental  mode $n=0$, i.e. to a fairly good
accuracy 
initial conditions (Eq.(\ref{5})) can be approximated
$B_y(\xi',t=0)= \alpha_0 e^{-{\xi'}^2/2 \sigma^2}$ (note that 
${\cal H}_0=1$). Therefore, substituting the latter expression
for $B_y(\xi',t=0)$ into the integral (\ref{6}), and evaluating the
integral analytically 
we obtain the solution of Eq.(\ref{4}) to the leading 
(fundamental $n=0$) order
\begin{eqnarray}
\nonumber
B_y={\frac{\alpha_0}{ \sqrt{1+ 
\eta C^\prime_A(x)^2 t^3/ 3 \sigma^2}}} \\
\times \exp{\left[{-{\frac{(z-C_A(x) t)^2}{2 (\sigma^2+
\eta C^\prime_A(x)^2 t^3/ 3)}}}\right]}, 
\label{8}
\end{eqnarray}
which coincides with the one obtained by \citet{hbw02}.
Here, $\sigma=0.033$ and $\alpha_0=1/5 \sqrt{2 \pi} \sigma$.
For large time, $t$, Eq.(\ref{8}) implies that the
amplitude of the Alfv\'en wave (e.g. amplitude of the $B_y$)
decays as 
\begin{equation}
B_y= \frac{1}{5} \left[2 \pi \eta C^\prime_A(x)^2/3\right]^{-1/2}\,t^{-3/2},
\label{9}
\end{equation}
i.e., in the case of a localized in $z$-coordinate 
Alfv\'en pulse, Heyvaerts \& Priest's exponential decay is
replaced by the power law $B_y \propto t^{-3/2}$ \citep{hbw02}.

Although, the both derivations in the end lead to
the same answer, in our opinion, the one presented here is more
rigorous, concise and it gives the 
general solution (Eq.(\ref{6})) and its asymptotic 
form (Eq.(\ref{6inf})) at once.

\end{document}